# Standard FTP and GridFTP protocols for international data transfer in Pamela Satellite Space Experiment

R. Esposito, P. Mastroserio, G. Tortone
*INFN, Napoli, I-80126, Italy*

F. M. Taurino
*INFM, Napoli, I-80126, Italy*

Physics applications often involve large amounts of data and/or computing and often require secure resource sharing across organizational boundaries, and are thus not easily handled by today's Internet and Web infrastructures. The Grid refers to an infrastructure that enables the integrated, collaborative use of high-end computers, networks, databases, and scientific instruments owned and managed by multiple organizations. The Globus project is developing fundamental technologies needed to build computational grids. Grids are persistent environments that enable software applications to integrate instruments, displays, computational and information resources that are managed by diverse organizations in widespread locations. Pamela (a Payload Antimatter Matter Exploration and Light-nuclei Astrophysics), is a permanent magnet core facility with a variety of specialized detectors. It allows the investigation of the cosmic radiation: origin and evolution of matter in the galaxy, search for antimatter and dark matter of cosmological significance, understanding of origin and acceleration of relativistic particles in the galaxy. Pamela will be put in an elliptical orbit at an altitude between 300 and 600 Km, on board of the Resurs DK1 Russian satellite in the fall of the year 2003. The Napoli INFN Computing Center and the Pamela Teams of the Universities of Rome, Trieste and Stockholm have performed several network tests in order to make a comparison between Standard FTP and GridFTP protocols. GridFTP is a high-performance, secure, reliable data transfer protocol optimized for high-bandwidth wide-area networks. It is based on FTP, the highly popular Internet file transfer protocol. Its main interesting features are: GSI security on control and data channels; Multiple data channels for parallel transfers; Partial file transfers; Third-party (direct server-to-server) transfers (it is possible, for example, that a user in Napoli can perform a file transfer from Stockholm to Trieste without getting involved); Resume interrupted data transfers. One result of our tests is reported as follow: a 1 GByte transfer file between the University of Stockholm and the University of Naples needed about 20 minutes using Standard Ftp; the same file was transferred in about 5 minutes using GridFTP. A discussion of the advantages of the configurations chosen is proposed below.

## 1. INTRODUCTION

**Grid** applications often involve large amounts of data and/or computing and often require secure resource sharing across organizational boundaries, and are thus not easily handled by today's Internet and Web infrastructures. Grids are persistent environments that enable software applications to integrate instruments, displays, computational and information resources that are managed by different organizations in widespread locations. Within this context the **Globus** project is developing fundamental technologies needed to build computational grids.

The **Globus Project** aims to develop new technologies that enable the creation of stable computational grids. Computational grids provide the computational infrastructure for powerful new tools for scientific investigation, including desktop supercomputing, smart instruments, collaborative environments, and distributed supercomputing. The strategy is to focus on services at the middleware level and generalize the requirements of grid applications and deliver technologies that support entire classes of applications. The middleware layer offers services for managing large numbers of diverse computational resources administered by independent organizations, and provides application developers with a simplified view of the resulting computational environment.

The Globus Project is engaged in defining and developing several capabilities necessary to build a persistent Data Grid environment: a **high-performance** and secure **data transfer mechanism**, a set of tools for creating and manipulating replicas of **large datasets** and a mechanism for maintaining a catalog of dataset replicas.

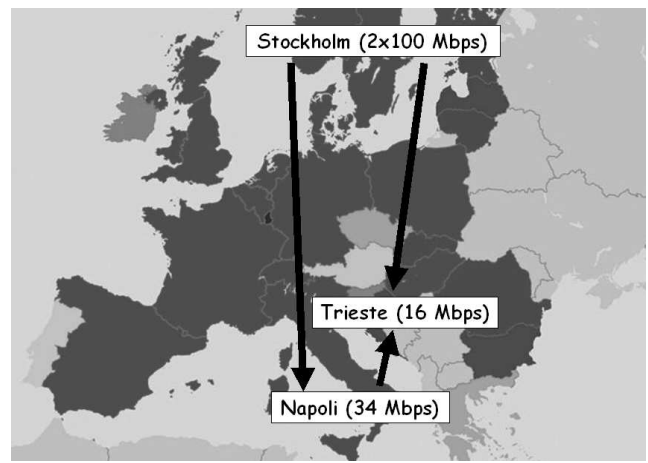

Figure 1: Connections to the Internet.

## 2. NETWORK TEST USING FTP

**FTP** is the most commonly used protocol for data transfer on the Internet, and the most likely candidate for meeting the grid's needs. It is attractive because is widely implemented and provides a well-defined architecture for protocol extensions. The **Pamela Satellite Space Experiment** has promoted same network test in order to decide which protocol to use in transferring files among its different sites in Europe and Russia. The first sites involved in these tests have been the Universities of Napoli, Trieste and Stockholm. A 1 GByte file has been transferred tens of times per night among the sites indicated above last year.

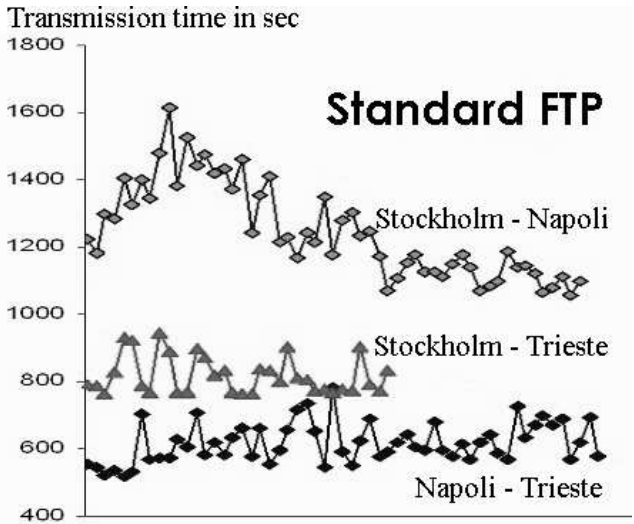

Figure 2: 1 GByte transfer file tests:
Stockholm – Napoli: 1268 sec average transmission time;
Stockholm – Trieste: 820 sec average transmission time;
Napoli – Trieste: 611 sec average transmission time.

The University of Stockholm is connected to the Internet with a 2x100 Mbps link, consequently the University of Trieste (16 Mbps link) and the University of Napoli (34 Mbps) are bottlenecks (fig.1). In fig. 2) the transmission times of 1 GByte file transfers are shown: The average time needed to transfer a 1 GByte file from Stockholm to Napoli has been of 1268 sec about, 820 sec about from Stockholm to Trieste and finally, 611 sec about from Napoli to Trieste. Interesting considerations can be done comparing the theoretical time needed to transfer a one gigabyte file from Stockholm to Napoli and the measured times in our tests.

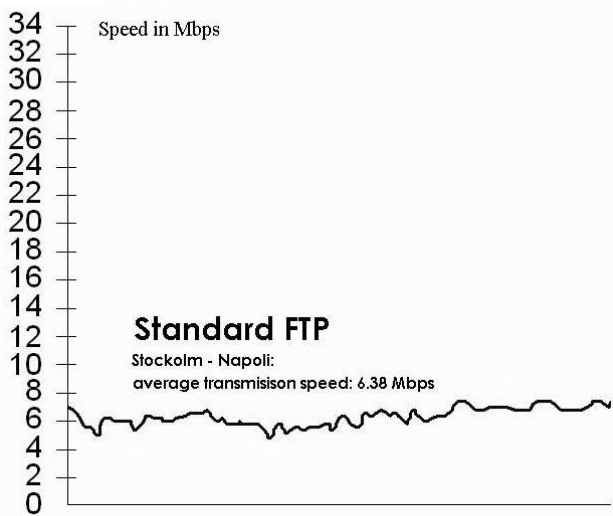

Figure 3: In Napoli, notwithstanding a 34 Mbps link to the Internet, the average speed of transmission tests has been of 6.38 Mbps.

In Napoli, as mentioned above, there is an ATM 34 Mbps link to the Internet; using TCP protocol on ATM (Asynchronous Transfer Mode), the maximum expected transmission speed is not more then 27÷28 Mbps due to the overhead of the protocols. In fig. 3) the same transmission tests are shown in terms of speed in Mbps: notwithstanding a 34 Mbps link to the Internet at our disposal in Napoli, the average speed of our transmission tests has been of 6.38 Mbps! It is necessary to look for a protocol that allows us to use all the bandwidth at our disposal.

## 3. NETWORK TEST USING GRID-FTP

**GridFTP** [1] is a selected subsets of the existing FTP protocol with new features; some of them are described below. GridFTP supports automatic negotiation of TCP buffer sizes both for large files and large sets of small files. On wide-area links, using multiple TCP streams can improve aggregate bandwidth over using a single TCP stream; GridFTP supports parallel data transfer. In order to manage large data sets it is necessary to provide third-party control of transfers between storage servers.

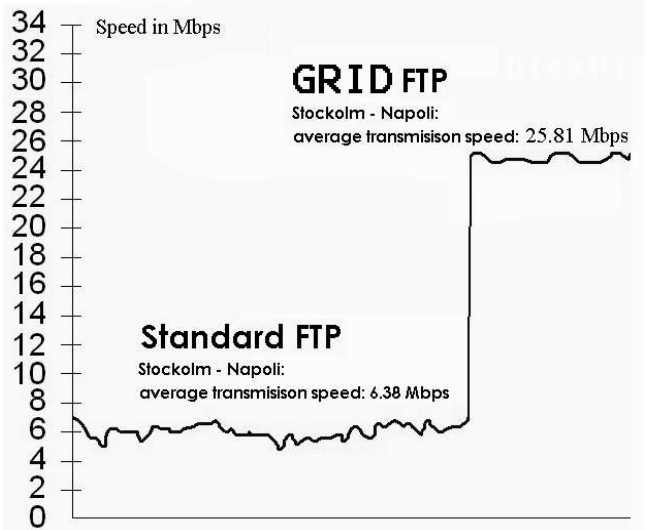

Figure 4: The average speed transferring files from Stockholm to Napoli has been of 25.81 Mbps using GridFTP.

GridFTP provides this capability by adding GSSAPI security to the existing third-party transfer capability defined in the FTP standard. Many applications require transfer of only a portion of a file. Transferring the entire file could be too expensive. GridFTP will support partial file transfer. The **Napoli INFN Computing Center** and the **Pamela Teams** of the **Universities of Trieste and Stockholm** have performed several network tests in order to make a comparison between **Standard FTP** and **GridFTP** protocols.

GridFTP is able to saturate the bandwidth at disposal, in fig. 4) the transmission speeds reached are shown; the average speed transferring 1 GByte files from Stockholm to Napoli has been of 25.81 Mbps, much more then the 6.38 Mbps using the Standard FTP.

## 4. CONCLUSIONS

An interesting consideration is given by the behavior of the two graphs shown in fig. 5): in the transmission times registered using GridFTP in our tests, the minimum and maximum transfer times differ no more then ten seconds from the average value as we have registered transmission times in a range that goes from 1100 to 1600 seconds using the Standard FTP. The good results obtained in these tests encourage the Pamela Computing Team to adopt GridFTP as standard tool for data distribution.

## References

[1] W. Allcock, J. Bester, J. Bresnahan, A. Chervenak, L. Liming, S. Meder, S. Tuecke, "GridFTP Protocol Specification", GGF GridFTP Working Group Document, September 2002.

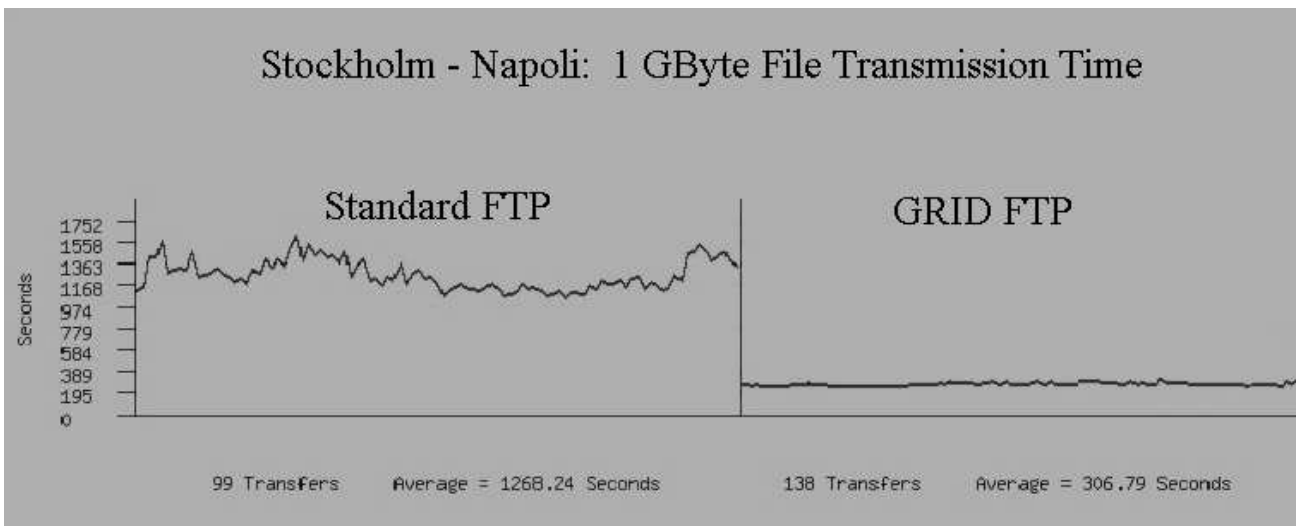

Figure 5: The transmission times in transferring 1 GByte file using Standard FTP and GridFTP are compared.

**TUCP008**